\documentclass{PoS}
\usepackage{subfigure}
\usepackage{amsmath}

\title{Calibration of the EUSO-TA detector with stars}

\ShortTitle{Calibration of the EUSO-TA detector with stars}

\author{\speaker{Z. Plebaniak}\thanks{for collaboration list see PoS(ICRC2019)1177}$^{a}$, J. Szabelski $^{a}$, M. Przybylak $^{a}$, L.W. Piotrowski $^{b}$, A. Djakonow $^{a,c}$, \newline K. Kr\'olik $^{a,d}$ for the JEM-EUSO Collaboration\\
        \llap{$^a$}National Centre for Nuclear Research, Astrophysics Division, Cosmic Ray Laboratory, \\ul. 28 Pu{\l }ku Strzelc\'ow Kaniowskich 69, 90-558 {\L }\'od\'z, Poland.
        \\
	\llap{$^b$} RIKEN, EUSO Team, Global Research Cluster, Wako, Japan \\
	\llap{$^c$} Faculty of Technical Physics, Information Technology and Applied Mathematics, {\L }\'od\'z University of Technology, ul. W\'olcza\'nska 215, 90-924, {\L }\'od\'z, Poland.\\
        \llap{$^d$} Faculty of Physics, University of Warsaw, ul. Pasteura 5, 02-093 Warsaw, Poland. \\
	E-mail: \email{zp@zpk.u.lodz.pl}	
        }

\abstract{The Extreme Universe Space Observatory-Telescope Array (EUSO-TA) is a ground-based experiment, part of the JEM-EUSO (Joint Experiment Missions - Extreme Universe Space Observatory) dedicated to the observation of Ultra High Energy Cosmic Rays (UHECRs) in parallel with the Telescope Array (TA) experiment. The main goal of EUSO-TA operations is to test the hardware and calibrate the EUSO detector to obtain optimal performance for cosmic ray observations. Apart from the artificial source calibration such as the Central Laser Facility (CLF), mobile lasers and UV diodes, natural signals from stars can be also used as a calibration source. This work presents the results of the calibration of the EUSO-TA detector. The influence of the atmosphere and of the detector parameters on star observations are discussed. Considering, stars as  point-like sources with well known UV emission parameters, signal amplitudes from stars as well as the EUSO-TA  detector point spread function were estimated. This unique calibration method could be used in future missions of the JEM-EUSO program such as EUSO-SPB2 (Super-Pressure Balloon).}

\FullConference{36th International Cosmic Ray Conference -ICRC2019-\\
		July 24th - August 1st, 2019\\
		Madison, WI, U.S.A.}

\begin{document}

\section{Introduction}

The Extreme Universe Space Observatory-Telescope Array (EUSO-TA) is a ground-based Extreme Universe Space Observatory (EUSO) family detector placed at the Black Rock Mesa site where one of the fluorescence stations of the Telescope Array (TA) \cite{TAMEDA200974} experiment is located.
This location allows for the measurements of Ultra High Energy Cosmic Rays (UHECR) in coincidence with the Telescope Array experiment. 


EUSO-TA is a Photo Detector Module (PDM) test for the next EUSO experiments, in operation since February 2015. During 5 observational campaigns in years 2015-1016, UHECR events  were detected and compared with simulations \cite{refId0}.
Measurements allow also for registration of others natural signals coming from stars, and meteors and artificial sources like the Central Laser Facility (CLF) or UV mobile laser provided by the  Colorado School of Mines group.


\begin{figure}
\centering
\includegraphics[width=0.7\textwidth]{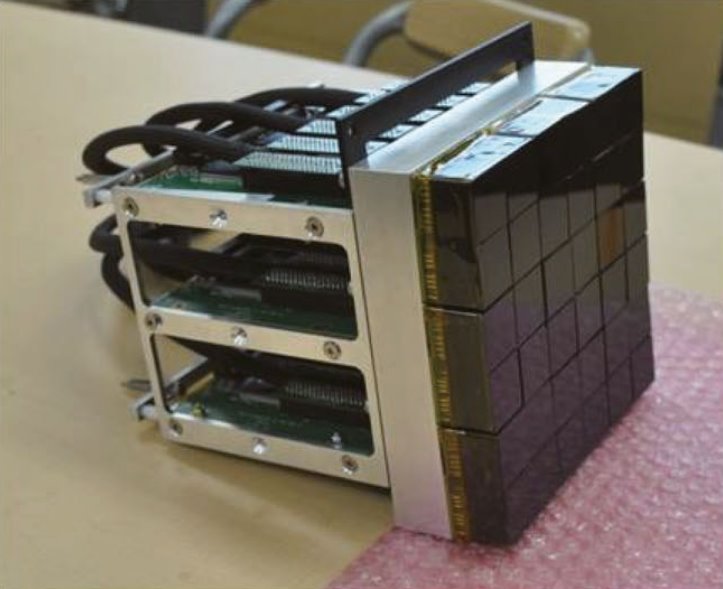}
\caption{PDM unit with array of MAPMT used in the EUSO-TA experiment \cite{Adams:2015pec}}
\label{PDM}
\end{figure}

The EUSO-TA telescope consists of two PMMA (polymethyl methacrylate) Fresnel lenses with effective areas equal to 0.92 m$^2$ \cite{EusoTaOptics} focusing photons on the focal surface where the PDM is placed.
The PDM is composed of 36 Hamamatsu R11265-M64 Multi-Anode Photomultipliers (MAPMT).
Four MAPMTs from Elementary Cell (EC).
Focal surface consists of 2304 pixels. The PDM design of EUSO-TA is similar to the ones used in both the EUSO-Balloon and  EUSO Super-Pressure Balloon 1 (EUSO-SPB1) \cite{Wiencke:2017cfi} experiments.

Observations of the night sky with a 10.6$^\circ\times$10.6$^\circ$ field of view (FOV) provides the opportunity to observe many passing stars through the detector focal surface.

The main idea described in this paper is to use signals that come from the stars to calibrate the EUSO-TA detector.

\section{Pixel equalization algorithm}
\label{equalization}

Pixel equalization of the EUSO-TA detector is necessary because pixels may have different efficiencies.
This operation allows for the consistent estimation of photon numbers registered from stars.
A common practice used for pixel equalization in case of imaging is flat fielding. This is realized by illuminating the telescope with a Lambertian light source.
In that case, we can estimate the relative response of each pixel for the same signal.
Unfortunately, this solution can not give us the global normalization (for whole PDM and for observation in the analyzed period) constant and the number of counted photons from the star will change as a function of the background level.
To solve this problem, we used experimental data, to estimate the efficiency of each pixel.
The basis of the pixel equalization method used here is presented below.
Measurements carried out in parallel with TA give us trigger rates of 4 Hz.
For each trigger we store one packet consisting of 128 frames (or GTUs - Gate Time Units). Each GTU corresponds to 2.5 $\mu$s hence as a result we have 320 $\mu$s of measurement per single packet.
During this time, the star almost does not change position on the focal surface.
This is why, for discussion purpose, we can integrate the counts in each packet.
Then, we can create the distribution of packet counts for each pixel.
If the background level during the short observation remains the same, counts distribution for one pixel can be described by a Gaussian function.

\begin{figure}
\includegraphics[width=0.51\textwidth]{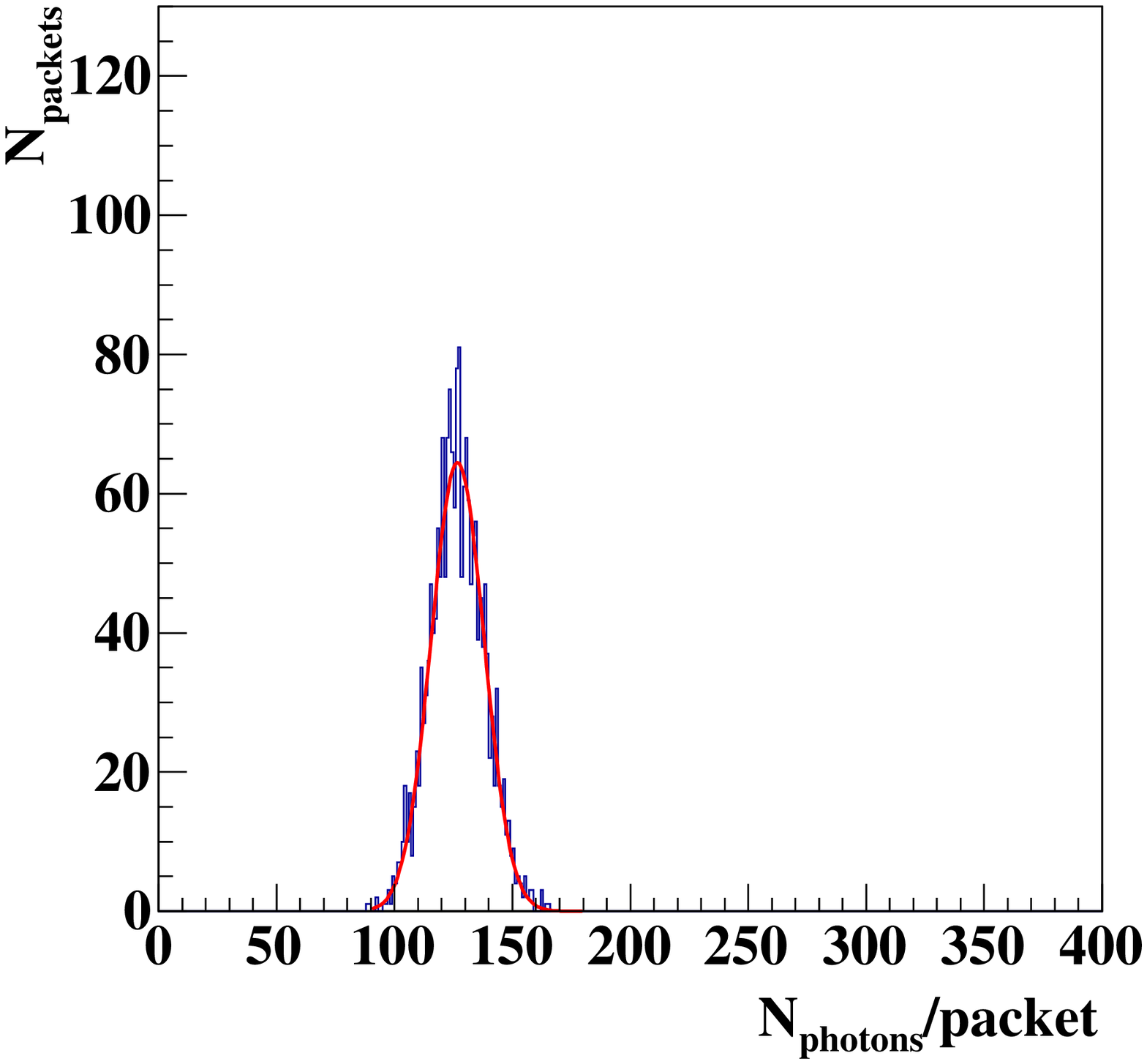}
\includegraphics[width=0.51\textwidth]{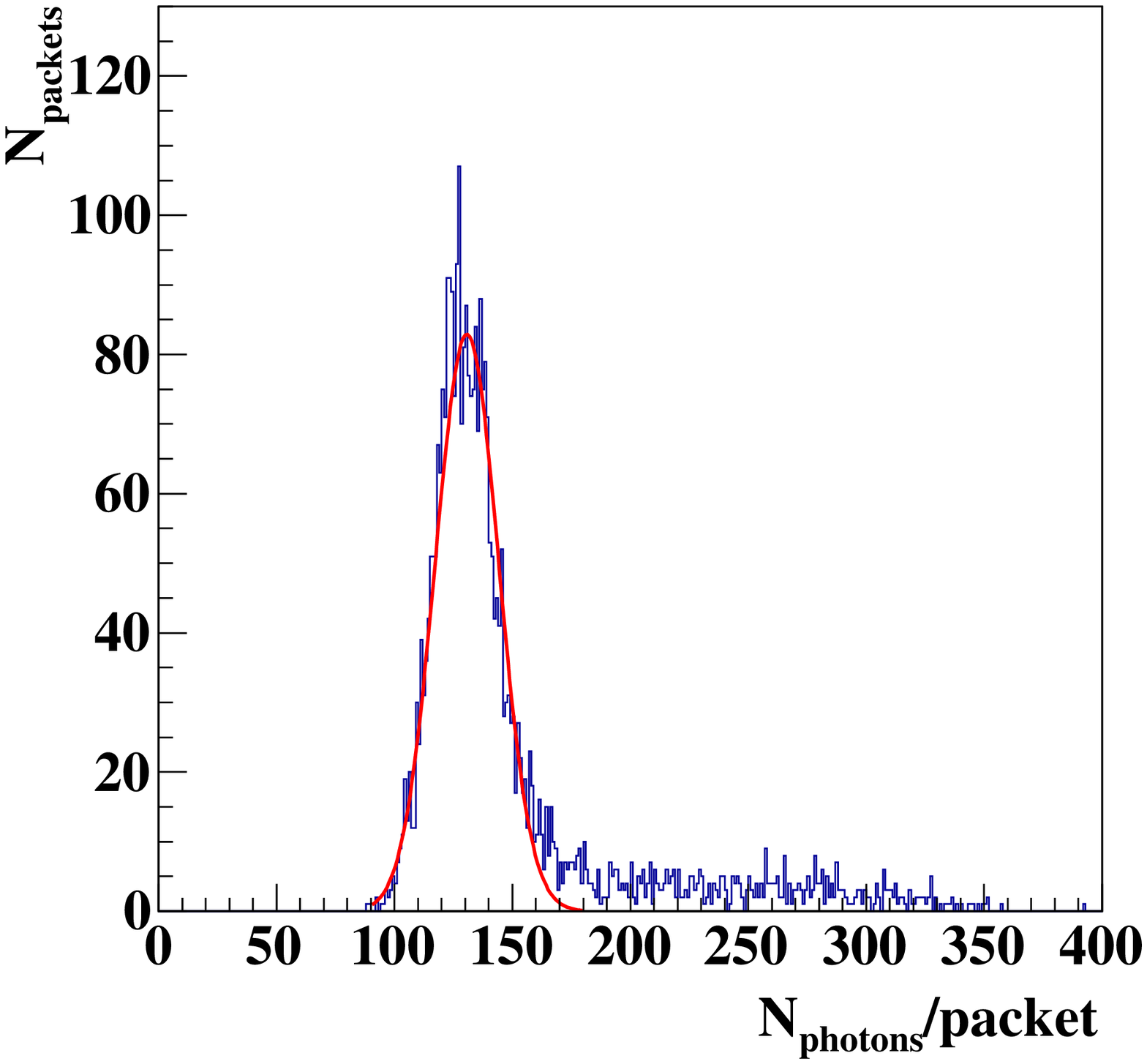}
\caption{Distributions of counts for one packet in one pixel in case of background only - 1000 packets (left) and additional higher signal due to a passage of a bright star through the pixel (right) - 3500 packets.}
\label{PixelDist}
\end{figure}

\begin{figure}[h]
\centering
\includegraphics[width=0.75\textwidth]{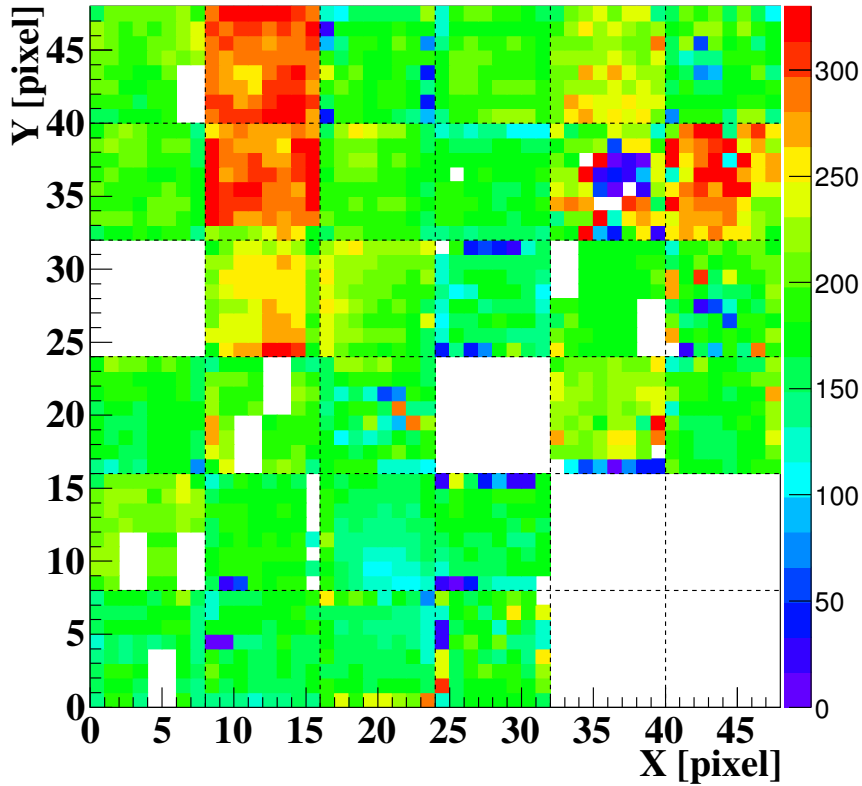}
\caption{Map of the pixels relative efficiency based on taken data. The values represent mean number of photons registered per packet.}
\label{PixelEffMap}
\end{figure}

An example of such kind of distributions in case of background only and with a signal of a bright star passing through the pixel  are analyzed in a pixel and is presented in Fig. \ref{PixelDist}.
In case of background only, the mean value of the Gaussian fit determines the relative efficiency of the analyzed pixel. The example of the relative efficiency map for all pixels on focal surface is shown in Fig. \ref{PixelEffMap}.
By creating a map of these values, we obtain results similar to the flat field case, but taking into account the background level.
To equalize the image related to one packet of data, we divide the signal for each pixel by the suitable efficiency and multiply by a normalization constant which is related to the average value of efficiencies for all the correctly-working pixels.
This can be described by the following formula:

\begin{equation}
SE_{x,y} = \frac{SR_{x,y}}{eff_{x,y}} \cdot A
\end{equation}
where:

$SE_{x,y}$ - equalized signal in {x,y} pixel

$SR_{x,y}$ - registered signal in {x,y} pixel

$eff_{x,y}$ - efficiency for {x,y} pixel (from Gaussian fit)

A - normalization defined by:

\begin{equation}
A = \frac{\left( \sum_{x = 0}^{48} \sum_{y=0}^{48} eff_{x,y} \right)}{N_{pix}}
\label{NormFactor}
\end{equation}
where $N_{pix}$ is the number of correctly working pixels in the analyzed area.
The above equation with x and y upper limits equal to 48 is related to the situation when we analyze the whole PDM surface with 2304 pixels.
This is the most simple case assuming that the background level in each pixel remains the same during observation.

EUSO-TA observes the sky from the ground level with an elevation angle ranging from 10 to 25 degrees. Part of the UV radiation passing through the atmosphere, both direct radiation from the stars and reflected from the surface of the Earth,is space absorbed, scattered  or reflected by the atmosphere. 
The intensity of these phenomena is closely related to the distance that the radiation overcomes in the atmosphere and with its density.
This causes the background to be different for pixels placed at  different elevations.
In that case, the normalization factor in Eq. \ref{NormFactor} can be calculated separately for each row of PMTs placed at different elevations.
Some stars with greater magnitude  can not be distinguished from the background. Signals from them are taken into account as a part of background.

\section{Stars in EUSO-TA data}
\label{Stars}
The discussed pixel equalization algorithm allows for fitting of signal position on the focal surface for stars with $M_{B}$ < 5.5$^{m}$.
In this magnitude limit, accuracy of the position fit is about 0.5 pixel using the fitting algorithm described in \cite{Plebaniak:2017iiu}.
Such bright stars can be visible as shown Fig. \ref{StarExample} (left).
Information about the magnitude of stars in the field of view comes from the HIPPARCOS catalogue \cite{Perryman:1997sa}.
Stars with lower brightness can not be distinguished from the background by eye, but up to $M_{B} \approx$  8.0$^{m}$ may be recognized in the light-curves for small pixel areas or separate photomultipliers in the way similar to plots in the Fig. \ref{StarExample} (right).

\begin{figure}
\includegraphics[width=0.50\textwidth]{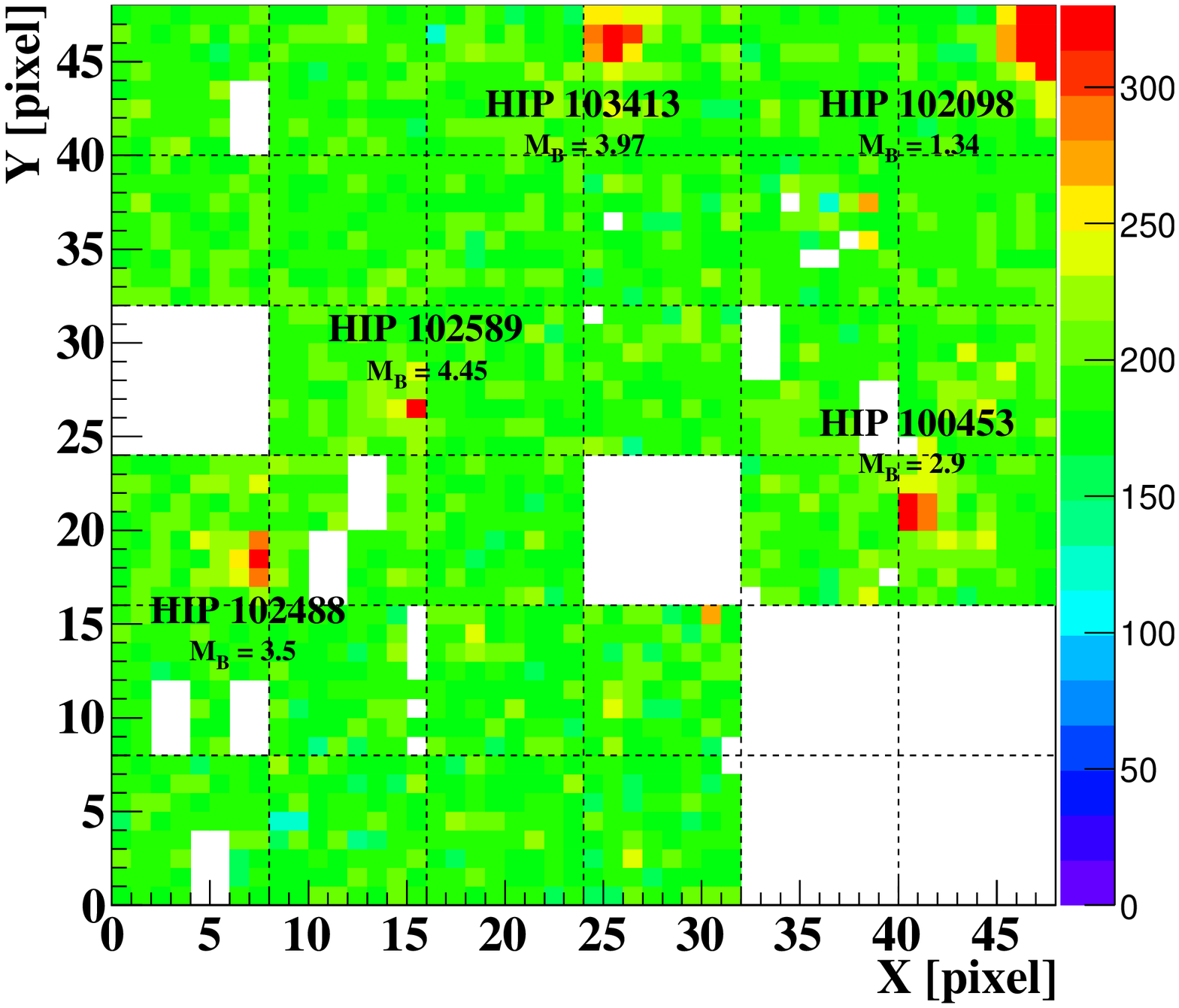}
\label{StarExample}
\includegraphics[width=0.52\textwidth]{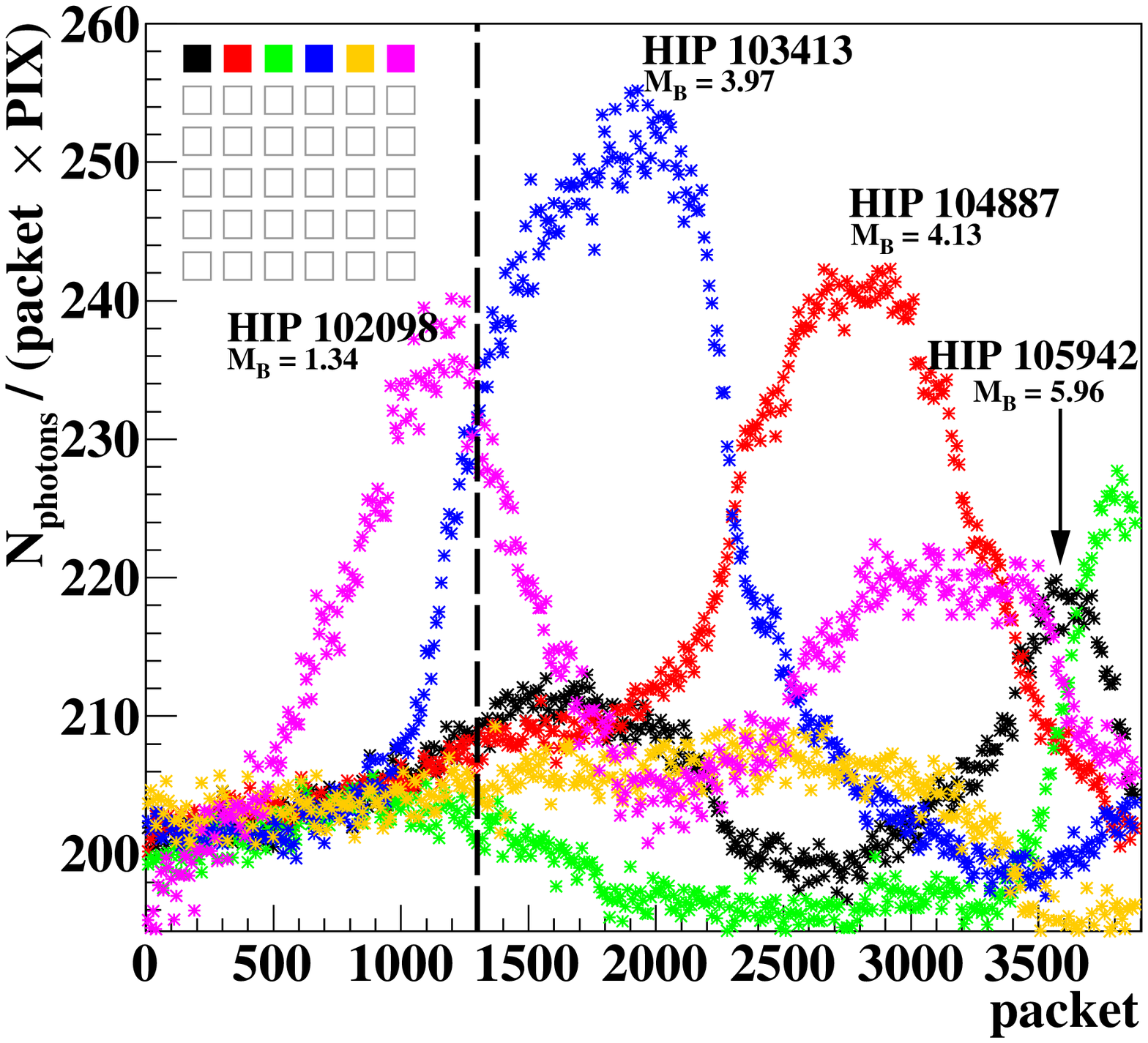}
\caption{Example of stars visible on the focal surface for one packet (left) and light-curves for the six upper MAPMTs (right) after pixel equalization. The vertical dashed line indicates the situation for the packet presented on the left picture.  }
\label{StarExample}
\end{figure}

\subsection{Measured Photons}
\begin{figure}[ht]
\includegraphics[width=0.54\textwidth]{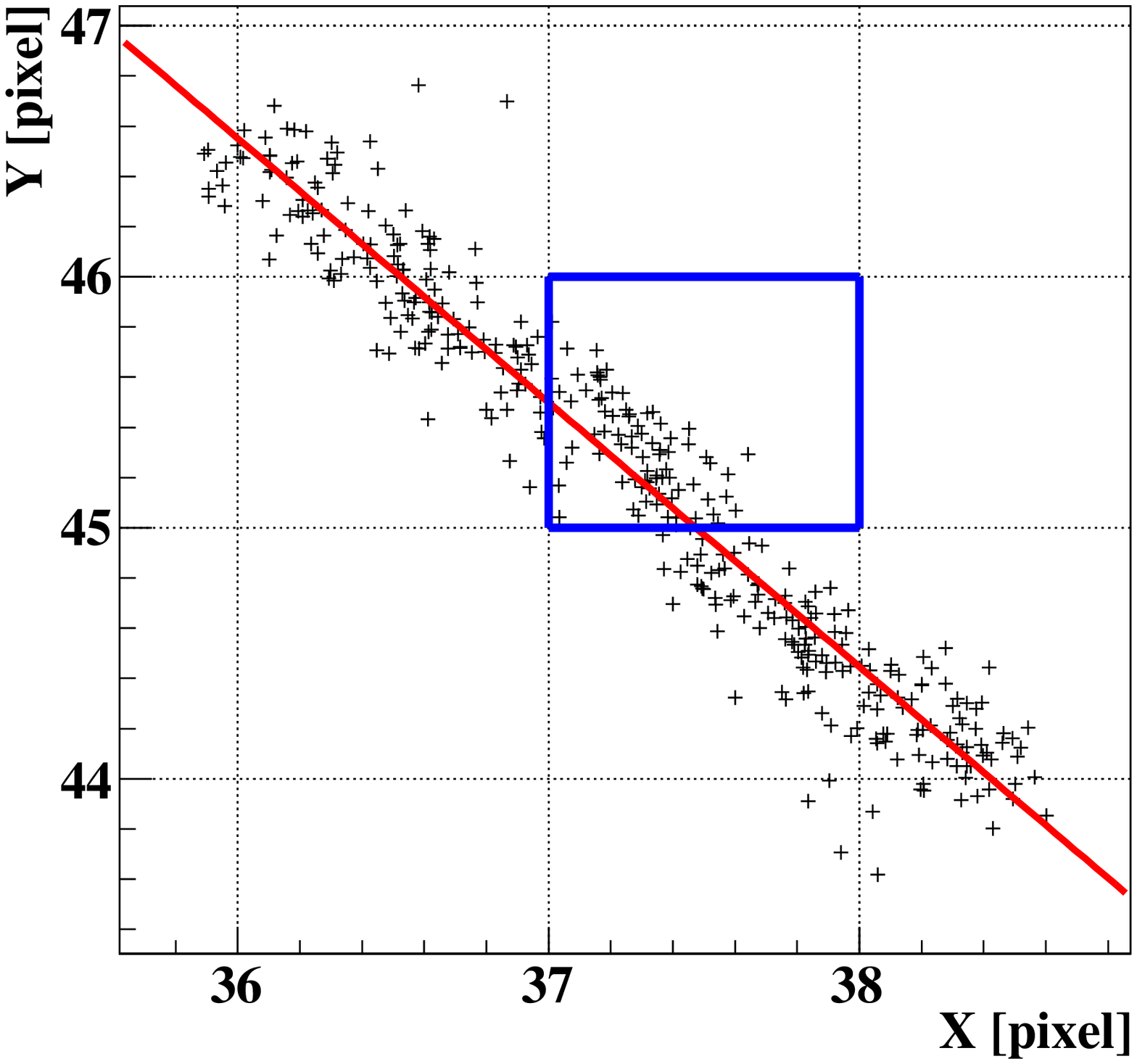}
\includegraphics[width=0.50\textwidth]{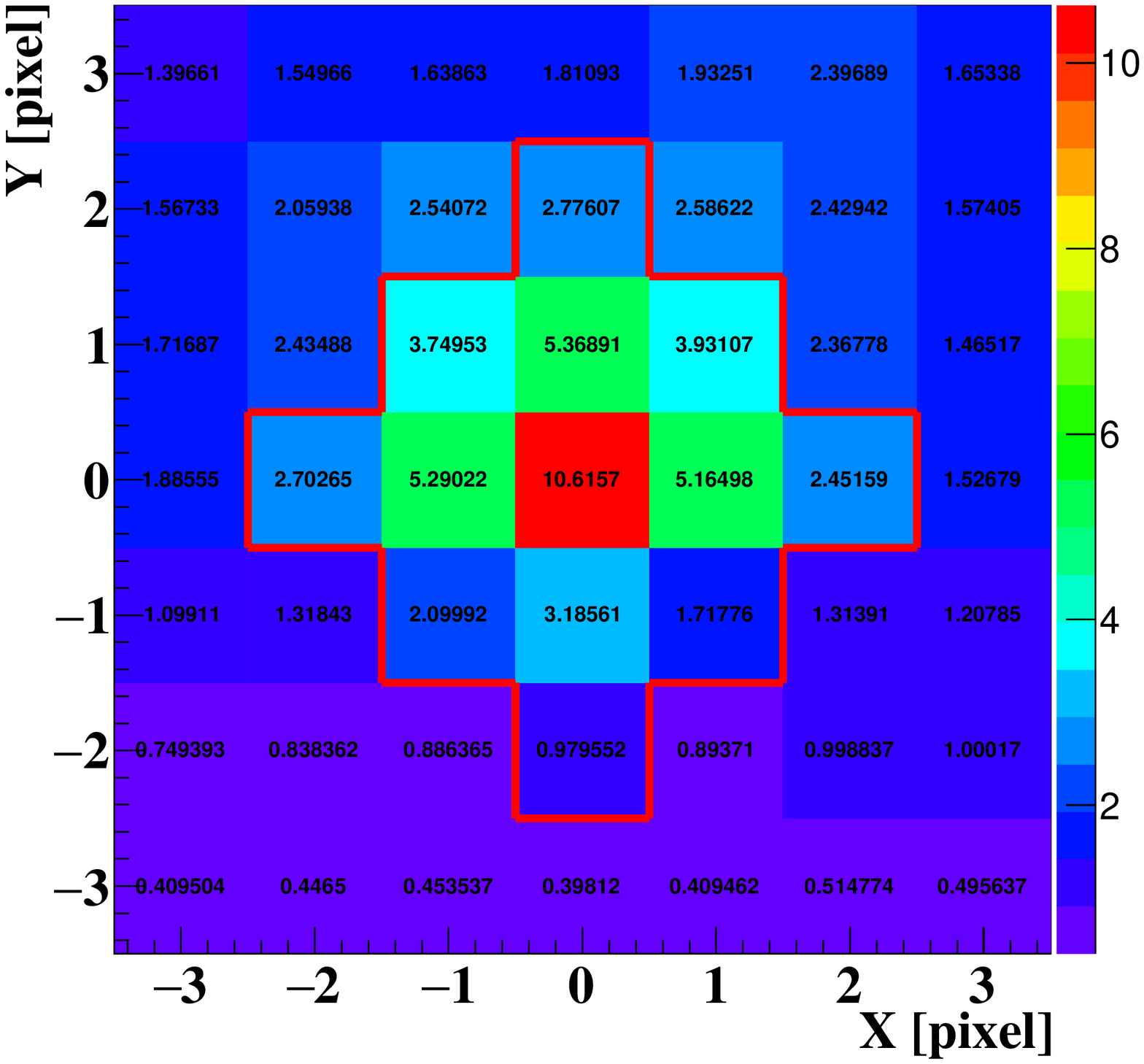}
\caption{Reconstructed star track on the focal surface (left) and percentage of signal from the star in different pixels in the selected area for photon counting (right).}
\label{StarsPhotonsAnalysis}
\end{figure}

The procedure dedicated to the estimation of the number of registered photons coming from stars is divided in few steps.
The first step is the described previously pixel equalization.
Then, the signal for a chosen star is fitted to determine the star track on the focal surface.
An example of the obtained track is presented in Fig. \ref{StarsPhotonsAnalysis} (left).
In the next step, we choose a pixel which star is traversing through, for example pixels marked by the blue rectangle in figure with track.
Then, for a long enough time, we calculate the number of registered photons per packet in area marked by the red line in Fig. \ref{StarsPhotonsAnalysis} (right), where the center of the cross is placed in previously chosen pixel.
As a result, we get a plot similar to the Fig. \ref{StarExample} (right) but related to the marked cross instead of all pixels in the MAPMT. The difference between minimum and maximum values represents the number of photons coming from star, while the minimum value indicates background level.
Photon counting within the cross forms instead of all PMT or even bigger area, allows to minimize the influence of other stars located near the analyzed one.
Results have uncertainties caused mostly by atmospheric conditions andinduce problems with star track fitting and variation of brightness during measurements.

\subsection{Expected number of photons}
\begin{figure}[ht]
\includegraphics[width=0.5\textwidth]{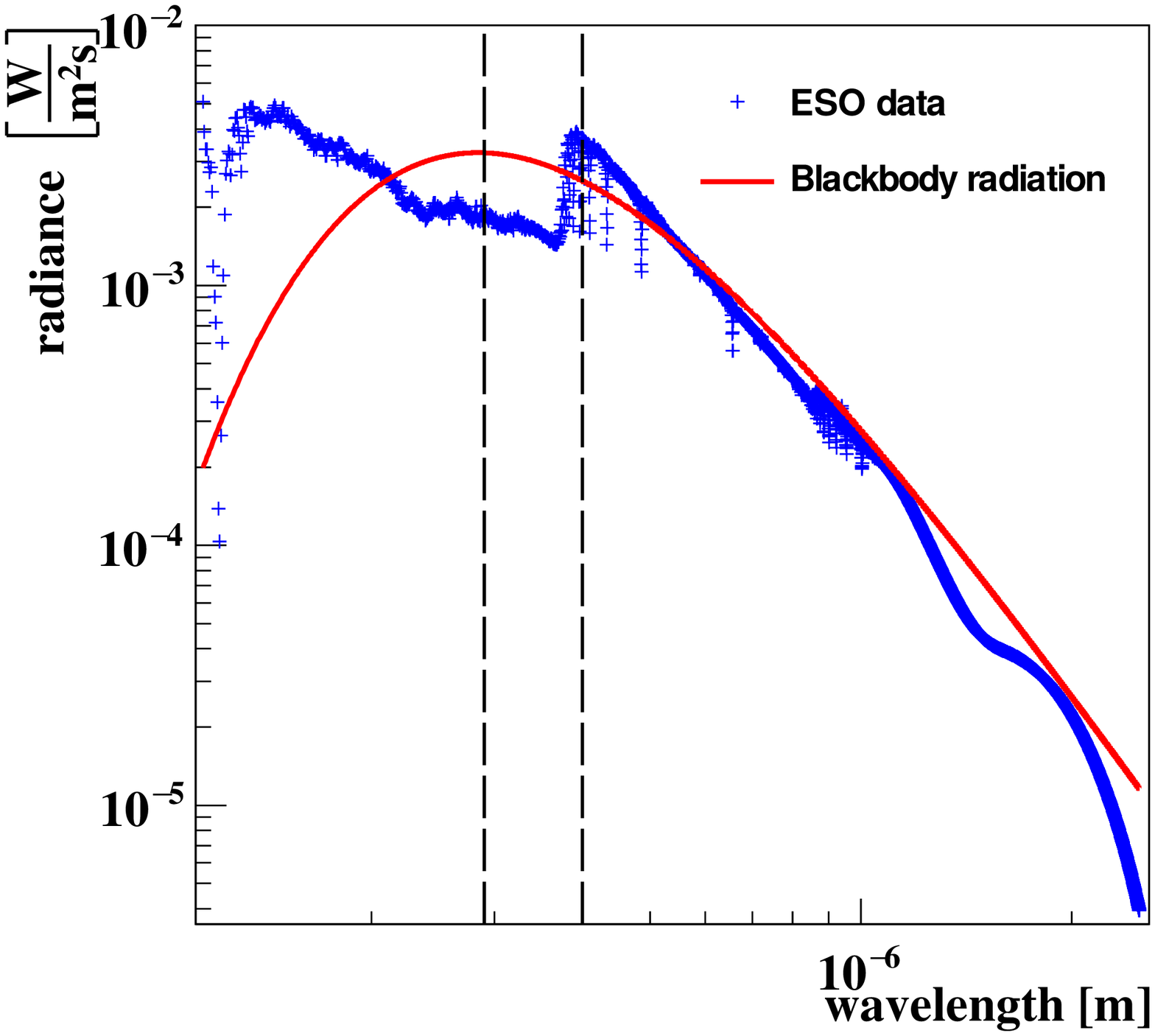}
\includegraphics[width=0.5\textwidth]{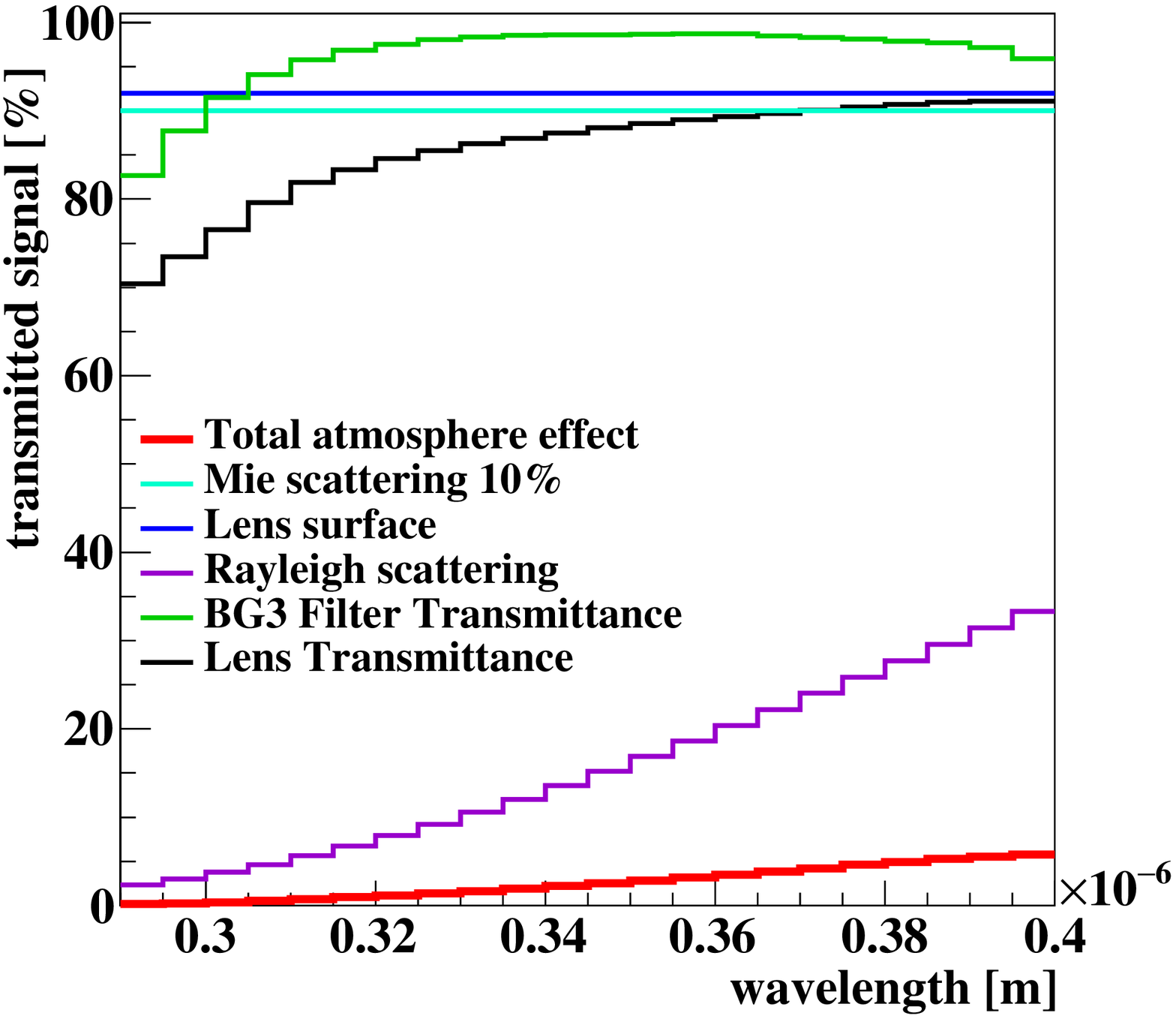}
\caption{Star spectra for spectral type B9III from ESO database, compared with related Planck distribution for star HIP111104 (left). Vertical lines in the left figure show EUSO wavelength range, and detector and atmospheric (scattering) parameters affecting the expected number of registered photons from star (right).}
\label{ExpectedPics}
\end{figure}

The estimation of expected signal is based on the stellar spectra data provided by the European Southern Observatory (ESO)  \cite{1998Msngr..94....7M}.
The necessity of using this kind of data is resulting from the fact that 
the stars behave according to blackbody radiation only as a first approximation.
An example of possible differences is presented in Fig. \ref{ExpectedPics} (left) for a star with spectral type B9III (e.g. HIP111104 in EUSO-TA data, see Tab. \ref{RegStars}).

Other effects should also be taken into account to estimate the number of expected photons correctly such as apparatus effects like quantum efficiency of MAPMT photocathode, transmittance of Fresnel lenses and of the BG3 filter.
Additionally, we use the Rayleigh and Mie scattering theory and US standard parameterization of the atmosphere to calculate the atmosphere opacity.
All those components as a function of wavelength in EUSO range are presented Fig. \ref{ExpectedPics}.
The red line represents the summarized effect as percentage of light amount reaching of the detector surface.

\begin{table}[h]
\centering
\caption{Results comparison for 7 measured stars. \label{RegStars}}
\centering
\resizebox{1.00\textwidth}{!}{\begin{minipage}{\textwidth}
\begin{tabular}{|c|c|c|c|c|c|}
\hline
HIP & M$_B$ & DATE & ANGLE & Measured N$_{ph}$/GTU & Expected N$_{ph}$/GTU \\
\hline
\hline
 83207 & 3.91 & 20161004-064103 & 11.0 &  2.88 &  3.77 \\
100453 & 2.90 & 20161002-100158 & 12.9 & 11.15 & 11.56 \\
 75411 & 4.60 & 20161006-041705 & 16.2 &  4.39 &  4.88 \\
111104 & 4.40 & 20161005-113107 & 17.5 &  6.24 & 11.98 \\
103413 & 3.96 & 20161002-100158 & 19.4 &  7.85 & 10.77 \\
111104 & 4.40 & 20161002-113051 & 19.5 &  6.53 & 15.18 \\
100453 & 2.90 & 20161005-091302 & 22.5 & 20.28 & 26.91 \\
\hline
\end{tabular}
\end{minipage}}
\end{table}

Calculated expected numbers of photons are compared with measurements for 7 examples of registered stars (see Tab. \ref{RegStars}).
The ANGLE column indicates the elevation angle related to the center of the MAPMT in which the star was observed.
Discrepancies between the observed and expected number of photons are probably caused in some part by different atmospheric conditions, but also by other factors which could be under investigation in future analysis.
According to the measurements of the used PMMA lenses \cite{EusoTaOptics}, the RMS spot size in case of EUSO-TA optical system is about 3$\times$3 pixels, which was taken into account.
Values in the last two columns of the table represent the number of photons per GTU  corresponding to photon counting per 2.5 $\mu$s.

\section{Inflight calibration of EUSO detectors}
\label{inflight}
The results of the measurements suggest that stars may be used as a calibration light sources during future EUSO missions.

Distributions in Fig. \ref{PixelDist}, produced with illumination of star light, can gives very useful information about the efficiency of each pixel on the focal surface.
In case of EUSO experiments operating in space or onboard stratospheric balloons such as EUSO-SPB2 \cite{Adams:2017fjh}, stars may be the only calibration sources.
In standard operation mode, EUSO-SPB2 will point to the ground and horizon direction from about 40 km altitude.
The idea for inflight calibration is to point the detector toward the sky to measure signals from stars, and use them calibration.
Even if this idea is not implemented during the EUSO-SPB2 mission, it could be implemented in other experiments.
Observations from above the atmosphere would give bright pictures of many stars, much better than during EUSO-TA observations from the ground.
Because experiments placed onboard stratospheric balloons may be rotating, it will be necessary to determine the pointing direction of the instrument in relation with the stars present its in FOV.

\section{Conclusions}

Based on the EUSO-TA data, we have developed the algorithm that allows to approximate the relative efficiencies of EUSO detector pixels, based on sky observations.
Thanks to the procedure described, we are able to equalize the recorded image of the sky in the way allowing for extracting of the signals from stars against the sky background. 
Estimation of the uncertainties of the registered number of photons require more extensive analysis related to the same star.
Additionally, the method allows to locate possible bad pixels during observations, and can be very useful for detector calibration in case of experiments located onboard stratospheric balloons or the space station.
Result of the presented analysis can contribute to the calibration of the EUSO detectors for the future missions.
\\

{\small{\bf Acknowledgments:} This work was supported by grant no 2015/19/N/ST9/03708 funded by National Science Centre in Poland.}



\bibliography{Bibliography}

\begin{thebibliography}{1}

\bibitem{TAMEDA200974}
Y.~Tameda.
\newblock Telescope array experiment.
\newblock {\em Nuclear Physics B - Proceedings Supplements}, 196:74 -- 79,
  2009.
\newblock Proceedings of the XV International Symposium on Very High Energy
  Cosmic Ray Interactions (ISVHECRI 2008).

\bibitem{refId0}
{Bisconti, F.}, {Belz, J.W.}, {Bertaina, M.E.}, {Casolino, M.}, {Ebisuzaki,
  T.}, {Eser, J.}, {Matthews, J.N.}, {Piotrowski, L.W.}, {Plebaniak, Z.},
  {Sagawa, H.}, {Sakaki, N.}, {Shin, H.}, {Shinozaki, K.}, {Sokolsky, P.},
  {Takizawa, Y.}, {Tameda, Y.}, and {Thomson, G.B.}
\newblock The detection of uhecrs with the euso-ta telescope.
\newblock {\em EPJ Web Conf.}, 210:05005, 2019.

\bibitem{Adams:2015pec}
J.~H. Adams.
\newblock {The EUSO-Balloon pathfinder}.
\newblock {\em Exper. Astron.}, 40(1):281--299, 2015.

\bibitem{EusoTaOptics}
Yoshiyuki Takizawa.
\newblock {The TA-EUSO and EUSO-Balloon optics designs}.
\newblock {\em PoS}, ICRC2013:0832.

\bibitem{Wiencke:2017cfi}
Lawrence Wiencke and Angela Olinto.
\newblock {EUSO-SPB1 Mission and Science}.
\newblock {\em PoS}, ICRC2017:1097, 2018.

\bibitem{Plebaniak:2017iiu}
Zbigniew Plebaniak, Jacek Szabelski, Tadeusz Wibig, and Lech Piotrowski.
\newblock {Point Spread Function of EUSO-TA detector}.
\newblock {\em PoS}, ICRC2017:460, 2018.

\bibitem{Perryman:1997sa}
M.~A.~C. Perryman et~al.
\newblock {The Hipparcos catalogue}.
\newblock {\em Astron. Astrophys.}, 323:L49--L52, 1997.

\bibitem{1998Msngr..94....7M}
A.~{Moorwood}, J.-G. {Cuby}, P.~{Biereichel}, J.~{Brynnel}, B.~{Delabre},
  N.~{Devillard}, A.~{van Dijsseldonk}, G.~{Finger}, H.~{Gemperlein},
  R.~{Gilmozzi}, T.~{Herlin}, G.~{Huster}, J.~{Knudstrup}, C.~{Lidman}, J.-L.
  {Lizon}, H.~{Mehrgan}, M.~{Meyer}, G.~{Nicolini}, M.~{Petr}, J.~{Spyromilio},
  and J.~{Stegmeier}.
\newblock {ISAAC sees first light at the VLT.}
\newblock {\em The Messenger}, 94:7--9, December 1998.

\bibitem{Adams:2017fjh}
James~H. Adams et~al.
\newblock {White paper on EUSO-SPB2}.
\newblock {\em arXiv:1703.04513}, arXiv:1703.04513, 2017.

\end{thebibliography}

\end{document}